\documentclass[preprint2]{aastex}

\shorttitle{Real-Time Analysis of Large Astronomical Images}
\shortauthors{Kuehn et al.}

\begin{document}

\title{Real-Time Analysis of Large Astronomical Images}


\author{K. Kuehn}
\affil{Argonne National Laboratory, Lemont, IL 60439}
\email{kkuehn@anl.gov}

\and

\author{R. Hupe}
\affil{Department of Physics, The Ohio State University, Columbus, OH 43210}

\begin{abstract}
Forthcoming instruments designed for high-cadence large-area surveys, such as the Dark Energy Survey and Large Synoptic Survey Telescope, will generate several GB of data products every few minutes during survey operations.  Since such surveys are designed to operate with minimal observer interaction, automated real-time analysis of these large images is necessary to ensure uninterrupted production of science-quality data.  We describe a software infrastructure suite designed to support such surveys, focusing particularly on ImageHealth, a tool for near-real-time processing of large images.  These image manipulation and analysis algorithms were applied to simulated data from the Dark Energy Survey, as well as observed data collected by the Y4KCam on the CTIO 1m telescope and the Mosaic camera on the Blanco telescope.  The accuracy and speed of the ImageHealth code in particular were benchmarked against results from SourceExtractor, a standard image analysis tool ubiquitous in the astronomical community.  ImageHealth is shown to provide comparable accuracy to SourceExtractor when examining bright objects in the focal plane, but with significantly shorter execution time.  Based on the importance of real-time analysis in reaching the Dark Energy Survey's science goals, ImageHealth and other aspects of this analysis package were incorporated (in modified form) into the Survey Image System Process Integration, the Dark Energy Camera software control environment.  The original ImageHealth code, however, is completely instrument-independent, and is freely available for use within other observational data-taking environments.
\end{abstract}


\section{The Need for Real-Time Analysis of Large Astronomical Datasets}
\begin{figure}
\epsscale{.75}
\plotone{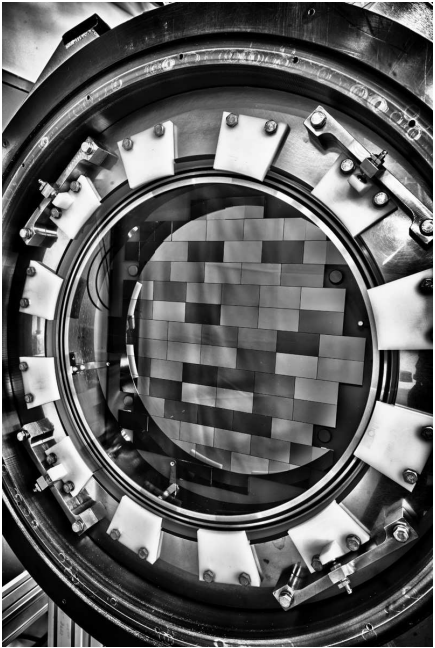}
\caption{A photograph of the 570 Megapixels of the DES focal plane prior to installation on the Blanco telescope at CTIO.}
\label{Fig1}
\end{figure}

Forthcoming instruments intended to be used for large-area high-cadence surveys such as the Dark Energy Survey, or DES \cite{Annis}, and the Large Synoptic Survey Telescope, or LSST \cite{LSST}, will potentially collect enormous amounts of data in an entirely automated fashion; therefore, observers need automated tools to analyze the performance of the instrument in real time.  The focal planes of these instruments are so large--DES images, for example, are of order 3 deg$^2$ and 1 GB in size (see Figs.~\ref{Fig1}), while LSST images will be a factor of several larger; furthermore, the cadence of these instruments is very rapid--DES images will be acquired approximately every two minutes over an entire 8- to 12-hour
night, while LSST images will be acquired at an even faster cadence.  Therefore, these tools must swiftly analyze a significant amount of information in a manner that facilitates immediate identification of problematic data and prompt recovery in the event of  observing malfunctions.  For example, if the telescope loses proper focus, all subsequent data taken by the survey instrument would be unusable for the key science analyses--especially the most sensitive ones, such as Weak Lensing.  Offline analysis of images by observers has no set cadence, so without an automated image quality analysis, significant observing resources could be wasted before human intervention identifies and corrects any problems.  In light of this need, a suite of software tools has been devised to fulfill the goal of rapid-response, ``quick and dirty'' image analysis.

\section{Image Analysis for the Dark Energy Survey and Beyond}
\begin{figure}
\epsscale{1.0}
\plotone{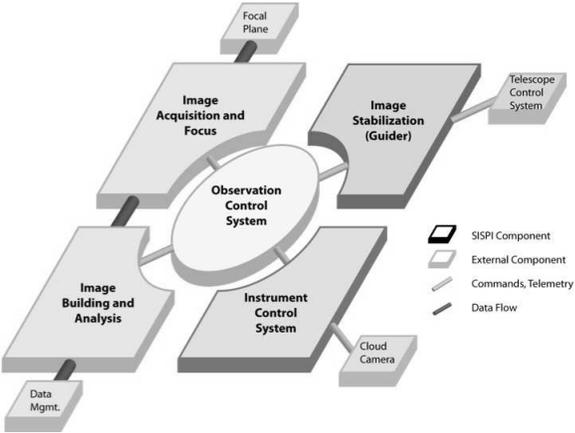}
\caption{A schematic of the SISPI components and the dataflow for DES data-taking.}
\label{Fig3}
\end{figure}

The ``central nervous system'' of the Dark Energy Camera (DECam), used for controlling all aspects of the instrument and providing images, telemetry, and other feedback to observers, is known as the Survey Image System Process Integration, or SISPI \cite{SISPI}.  Figure \ref{Fig3} shows a schematic representation of the SISPI components and the flow of data.  Several independent image manipulation and analysis modules have been incorporated into the Observer Control and Instrument Control aspects of SISPI in order to provide near-real-time information to observers.  These tools include a Real Time Display, which provides a static, compressed (from 1GB to 4MB) focal plane image within 1s; Instrument Health, which accumulates telemetry information and prepares time-history plots of data important to maintaining the integrity of the observing process; Quick Reduce, which performs more sophisticated image processing (e.g., astrometry and photometry) on a selection of images within the dataflow, where depth of analysis is more important than immediate feedback; and ImageHealth, which performs a very fast check of CCDs similar to the widely-used Source Extractor \cite{SEx}, but more streamlined in execution and output.  In addition to all of the tools present within the camera infrastructure, separate ``off-line'' software has been developed for the full analysis of all data obtained for the Dark Energy Survey.

The various software subsystems within SISPI will be described in forthcoming publications of the DES Collaboration; in the rest of this work, we focus on the specific software tool designed for the broadest applicability outside of the DES: the ImageHealth algorithm.  Section 3 describes the steps of the algorithm, while Section 4 describes the variety of user-defined input parameters that can be modified for algorithm performance optimization in a wide variety of observational settings.  Section 5 provides instructions on usage and the describes format of the code output, Section 6 provides a quantitative analysis of ImageHealth results, and Section 7 provides the conclusions drawn from this analysis.

\section{The Image Health Algorithm}
ImageHealth (IH) is designed to rapidly identify relatively bright objects within an image, and then swiftly determine a few specific image- and object-based parameters that can diagnose the overall quality of the image.  While the DES-specific version has undergone modification for integration into the Survey's software infrastructure, in its most basic, instrument-independent form ImageHealth is a C program that incorporates the standard CFITSIO library \cite{CFITS}, and executes the following steps:
\begin{itemize}
\item
Opens image from a file, receives relevant FITS image header keyword values (e.g., saturation, axis size), and determines the number of extensions contained in the file.
\item
Performs the following over all the image extensions two times--once for the right half and once for the left half of each image extension:
\begin{itemize}
\item
Reads pixel values to a data structure.
\item
Finds mean of (non-saturated, non-dead) pixel values, including either every pixel or every N$^{th}$ pixel (as determined by the ``pixel skip'' value).
\item
Determines location of a bright pixel, which becomes the first (seed) pixel of the ``Object''.
\item
Finds all pixels associated with that Object.  If any associated pixels are saturated, a new Object is selected and the old Object is discarded.
\item
Calculates Object (and full image) statistics:  Background counts near Object and Mean Sky Background counts over half of the image extension; Number of Dead, Saturated, and Good pixels; Object Flux (in counts); Object Size (FWHM, in pixels); Object Major and Minor Axis (in pixels); and Object Ellipticity and Orientation Angle of Object ellipse.  This narrowly-defined set of parameters provides the basis for effectively diagnosing some of the most common scenarios from which out-of-tolerance observations will arise, including: voltage or thermal control failures in the instrumentation, sub-optimal focus settings, optical element mis-alignment, and poor atmospheric seeing (among others).
\end{itemize}
\item
Finally, IH outputs quantities for Image-by-Image and Time History Displays.  In particular, anomalous quantities can be tagged to alert the user that intervention in the automated observing process is required.
\end{itemize}

Sometimes objects do not exist in a given image region, or they are too faint for an algorithm geared toward the detection of bright objects.  Though in general 100$\%$ coverage is not achieved by IH (that is, a suitable Object is not always identified in every FITS extension), the number of Objects found is clearly sufficient to broadly characterize the overall performance of the instrument and the general quality of any given image. For example, DES images are recorded by 124 separate readouts (2 for each of 62 CCDs/FITS extensions); out of the 124 distinct iterations of the IH algorithm applied to these image subregions, $\approx$100 Objects (on at least separate 50 CCDs covering 80$\%$ of the focal plane) are routinely discovered, with the exact value being mildly filter- and threshold-dependent.  

\section{User-Defined Parameters of the Image Health Algorithm}
Prior to compiling and executing \tt{imagehealth.c}\rm, the user can define several parameters within the code.  Some of these parameters are instrument-specific and are not likely to be changed otherwise, such as:
\begin{itemize}
\item
\tt{OVERSCAN}\rm: Size of the overscan region (in pixels) on each chip; this region is ignored by the algorithm.
\item
\tt{DEAD}\rm: Minimum pixel value (in ADU); pixels below this value are considered dead and are not incorporated into any calculations performed by the code.
\item
\tt{SATVAL}\rm: Pixel Saturation Value (in ADU); pixels above this value are not incorporated into any calculations performed by the code, and Objects with saturated pixels are ignored by the algorithm.  Setting this parameter to 0 causes the algorithm to read the saturation value from the FITS Header.
\item
\tt{MINPIX}\rm: Minimum Object Size (in pixels); objects containing fewer pixels above a certain limit (see below) than this number are ignored by the algorithm.  Setting this parameter to 6 or more (depending on the plate scale) provides effective rejection of potential cosmic rays without eliminating a significant number of bright Objects useful to this algorithm.
\item
\tt{NUMPIX}\rm: Object Searchbin Size (in pixels) determines 1/2 of the side length of the rectangular search area for pixels that may be considered part of the Object.  The search area is centered on the Object Seed pixel.  This parameter is commonly set to 12, which will completely contain most Objects (depending on the plate scale), unless the instrument focus is egregiously bad or the Object itself is widely extended across the sky.
\end{itemize}

Other parameters are likely to be changed more regularly by users, depending on their specific goals and the nature of the data that they are analyzing.  These include:
\begin{itemize}
\item
\tt{NMOVE}\rm: The number of FITS extensions incremented before each subsequent iteration of the code.  Unless the user has a reason to skip the analysis of some portion of an image, this should be set to 1--i.e. each subsequent iteration of the code acts on (current extension + 1).  If there is only a single extension to the FITS file, regardless of the value of NMOVE the algorithm is executed on that extension and then ImageHealth closes normally (as it does whenever it reaches the end of a FITS file).
\item
\tt{THRESHOLD}\rm: Object Seed Threshold, commonly set to 10, determines the multiplicative factor above the mean sky background that a pixel must have in order for it to serve as the seed for an Object.  Lower values such as 5 or even 3 can be used if the field is populated only by fainter objects that do not result in sufficient Objects for analysis of the complete focal plane.
\item
\tt{OBJTHRESH}\rm: Object Pixel Threshold, the value in sigma above the image mean required for any pixel to be considered part of an Object (rather than a Background pixel).  A pixel threshold of 5 results in mean photometric accuracy of $\approx$0.5\% (see Section 6) for the bright objects that this algorithm is designed to find.  Fainter objects or those with broad wings are likely to be less accurately measured, as there could be several pixels that in reality should be considered part of the Object, but are below the default parameter value.  In cases where many such Objects are found, a smaller value for this parameter (like 3) may be more appropriate.
\item
\tt{PIXSTEP}\rm: ``Full Counting'' or ``Pixel Skipping'' determines whether or not every pixel is read into the data structure.  Common pixel skip values are 2, 4, or 8, which allows the algorithm to execute faster with minimal sacrifice of accuracy.  The default value of 1 results in every pixel being counted.
\item
\tt{NLOOPS}\rm: Limit to Number of Potential Objects which may be found before the algorithm proceeds to the next iteration.  If this number is exceeded, the algorithm's current iteration is terminated and it begins to examine the next subregion or FITS extension.  This is useful for constraining the computing time expended on an image extension which may have many objects, but few which would be useful for the IH algorithm.  Combined with the thresholds defined above, an \tt{NLOOPS}\rm~of 6 results in Objects found in a high fraction of all extensions for a wide variety of test images.
\end{itemize}

\begin{figure}
\epsscale{1.0}
\plotone{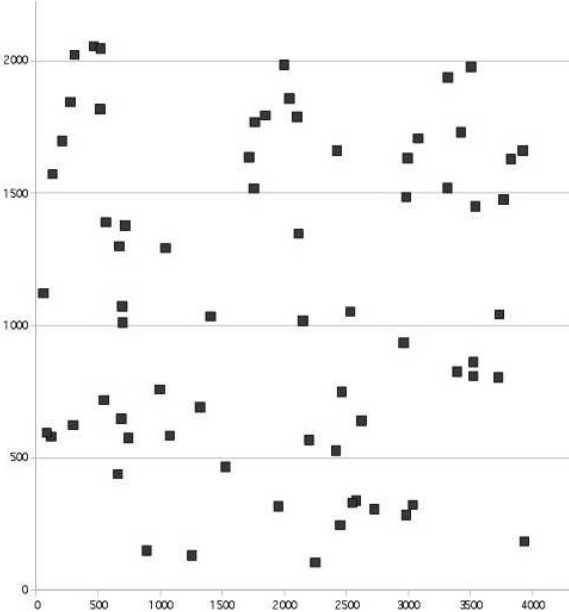}
\caption{X and Y positions of Objects found by the ImageHealth Algorithm on 2048x4096 DES CCDs.}
\label{Fig5}
\end{figure}

While the constant starting pixel position of the search algorithm could potentially introduce a bias in the image location of Objects found by IH (e.g., all Objects could be clustered in one corner of each CCD), in practice the Objects are relatively evenly distributed throughout the CCDs--see Fig.~\ref{Fig5} for a representative sample of Object positions.  If users discover unacceptably nonuniform distributions of Objects, then setting \tt{THRESHOLD}\rm, \tt{NLOOPS}\rm, or \tt{PIXSTEP}\rm~to higher values will likely result in different (and more uniform) Object selection.  Source crowding or image seeing may also impact the choice of parameter values, especially \tt{PIXSTEP}\rm~(or perhaps \tt{NUMPIX}\rm), though optimal parameter values under non-standard observing conditions are best determined by users on a case-by-case basis.

\section{Using Image Health}
Once the user has set the desired values for all parameters via the \tt{\#define}\rm~statements within \tt{imagehealth.c}\rm, the program is compiled with (for example) gcc:\\
\tt{\% gcc -lm -lcfitsio -L/CFITSIO/PATH -o ih}\rm\\
and then executed from the command line:\\
\tt{\% ./ih inputfile.fits outputfile.dat}\rm\\
If the syntax is not followed appropriately, an error message will be generated describing the proper syntax.  Note that the compile step requires the CFITSIO libraries to be properly installed (\tt{/CFITSIO/PATH}\rm~ should also be altered to reflect the user's actual directory structure).  The execute step further assumes that \tt{inputfile.fits}\rm~is in the directory where the code is executed; if it is not, then the appropriate path and filename should be specified by the user.

By default during execution, a variety of status, warning, or error messages are printed to the specified output file, though if user does not wish to see them they are trivial to comment out.  Likewise, the quantities calculated for each half or each FITS extension (number of dead or saturated or good pixels, sky background) as well as the quantities for the Object (X/Y position, local background, flux, FWHM, major and minor axis lengths, ellipticity and orientation angle) found on each half of each extension are printed to the specified output file.  At the conclusion of the execution, the mean FWHM of all Objects is returned.  This single number provides a useful summary of the characteristics of the entire image, though the user will likely want to examine the output file in detail to determine the true health of the image and its constituent extensions.

In the next section we describe the comparison of the Image Health algorithm to that of another common image analysis tool, Source Extractor.

\section{Comparison of ImageHealth with Source Extractor}

Source Extractor (SE) is a standard image analysis and object-finding software
package used for a wide variety of tasks by many in the astronomical community, and it also serves as our benchmark for comparison to the results
of the ImageHealth algorithm.  Analysis of hundreds of DES simulated images \cite{DesSim} with both ImageHealth (using the default settings intended primarily for the evaluation of bright Objects) and Source Extractor show very good agreement in their outputs; see Table~\ref{Tab1} for details.  The supplemental code (beyond \tt{imagehealth.c}\rm) that performed these large-scale comparisons between the two packages can be provided upon request from the authors.

While IH does not have the versatility of SE, it does exhibit comparable performance for the specific parameters calculated, provided there are sufficient numbers of bright stars in each image. Additionally, the handful of input parameters and single mode of execution gives IH not only greater user-friendliness than SE, but also facilitates greater modification of the algorithm and its outputs by any and all users.  ImageHealth is comprised of only 600 lines of code, while the numerous different aspects of Source Extractor total 1700 lines.  Furthermore, IH is written modularly, with distinct and clearly
delineated operations for file I/O and object analysis--processes which, in Source Extractor, are intertwined in a
matter not at all transparent to the user.  Finally, IH offers streamlined processing with more focused
output: for those applications where execution time is a significant factor, it is worth noting that on a standard (2.2 GHz) desktop machine, IH executes in 30 seconds in full counting mode, or as little as six seconds in pixel-skipping mode.  On the same machine, Source Extractor takes 70 seconds to step through a
full ($\approx$1GB) DES image.  In the context of the Dark Energy Survey observing cadence (images acquired every two minutes or less), IH 
will determine the quality of image N--and allow significant time for observer intervention--well before image N+1 is read out.  Thus, in the event of errors in the 
observation or problems with the data, at most a single image (and two minutes of observing time) is lost.  The time differential (up to a factor of 10) between Source Extractor and IH execution may be even more crucial for Community Users of the DECam, who will have a wide variety of observational requirements, including, perhaps, even faster cadences.  Similarly, LSST, with individual images several times larger than DES images and observations occurring at a faster cadence than DES, could benefit from implementing ImageHealth over Source Extractor for real-time image analysis solutions where only a few select output parameters are required.

\begin{table*}
\centering
\begin{tabular}{lrrr}
\hline
  Parameter
  & Mean IH-SE
  & Mean IH-SE (Pixel Skip = 8)
  & Units   \\
\hline
Position & 1.8 & 1.72 & pixels \\
Object BG & 0.35 & 0.24 & \% of counts \\
Object Flux & 0.53 & 0.68 & \% of counts \\
FWHM & 0.22 & 0.24 & pixels \\
Ellipticity & 0.03  & 0.04  & ...\\
Major Axis Orientation Angle & 8.6 & 7.4 & degrees \\
\hline
\end{tabular}
\caption{Sample Results of the ImageHealth
algorithm and identical parameters determined by Source Extractor (SE)}
\label{Tab1}
\end{table*}

While the ImageHealth algorithm has been tested on many (simulated) DES images, it is not instrument-specific.  The algorithm was also tested on Y4KCam \cite{Y4KCam} images, both from science and engineering observing periods, and performed with comparable accuracy and speed to the tests on simulated DES images.  Since the DES and Y4KCam datasets primarily comprised simulated or observed ``good'' images, additional tests were performed on images of the Coma Cluster taken with the Mosaic camera on the Blanco telescope that were known to be of poor quality.  While the former tests show that ``false positive'' mis-identification of problems is avoided by IH, the most important result from this final test is that ``false negative'' mis-identification of truly problematic images is likewise avoided.  Specifically, low quality, large FWHM images were readily and routinely identified as such by the ImageHealth algorithm.  Subsequent to the completion of these formal tests, ImageHealth was also explored as a source of real-time feedback for the MODS instrument \cite{MODS}.  There are even image processing applications outside the field of astronomy (e.g., in mechanical and electrical engineering) for which the swift and accurate processing of large datasets that ImageHealth offers is proposed as the most appropriate solution \cite{Monon}.

\section{Conclusions}

While Source Extractor is a broadly useful tool for the astronomical community, ImageHealth is shown to be the superior application specifically for observers who require real-time feedback on a well-defined set of parameters useful for determining image quality and instrument performance.  This code has been modified for use within the Dark Energy Survey's Survey Image System Process Integration, the software infrastructure that will be used by the Dark Energy Survey and all Community Users of DECam.  Furthermore, the instrument-independent version of the code is freely available from the Astrophysics Source Code Library \cite{ASCL} to all astronomers desiring to perform real-time analysis of large-scale observations using completely different instruments.

\acknowledgments
The authors wish to thank Professor Rick Pogge of The Ohio State University Department of Astronomy for useful feedback on early stages of the project, as well as access to archival Y4KCam data used in the testing of the ImageHealth algorithm.  Thanks as well to Professor Klaus Honscheid for his support of these efforts, and to Professor Alex Small and Dr. Lisa Gerhardt for productive comments on early drafts of this work.

\clearpage




\begin{thebibliography}{}
\bibitem[Abell et al. 2009]{LSST}
Abell, A., et al., LSST Science Book, V. 2.0; available online at arXiv:astro-ph/0912.0201

\bibitem[Allen et al. 2012]{ASCL}
Allen, A., et al., \emph{American Astronomical Society Annual Meeting 219, $\#$145.10}

The ImageHealth code and supplemental files are available from the Astrophysics Source Code Library at:
http://asterisk.apod.com/viewtopic.php?f=35\&t=28890

\bibitem[Annis et al. 2005]{Annis} Annis, J. et al., White Paper Submitted to the Dark Energy Task Force; available online at arXiv:astro-ph/0510195 (2005)

For more information see http://www.darkenergysurvey.org

\bibitem[Bertin $\&$ Arnouts 1996]{SEx}
Bertin, E. and Arnouts, S., \emph{Astronomy $\&$ Astrophysics Supplement} 317 (1996) 393

\bibitem[Honscheid et al. 2008]{SISPI}
Honscheid, K., et al., \emph{Proceedings of SPIE} 7019 701911 (2008)

\bibitem[Kuropatkin et al. 2012]{DesSim}
Kuropatkin, N., et al., in prep.
 
\bibitem[Mahboob 2009]{Monon}  
Mahboob, M., private communication

\bibitem[Osmer et al. 2000]{MODS}
Osmer, P.~S., et al., \emph{Society of Photo-Optical Instrumentation Engineers Proceedings} 
4008 (2000) 40

\bibitem[Pence et al. 1999]{CFITS}
Pence, W., et al., \emph{Astronomical Society of the Pacific Conference Series} 172 (1999) 487

For more information and downloadable software, see 
http://heasarc.nasa.gov/fitsio/fitsio.html

\bibitem[Pogge 2009]{Y4KCam}
Pogge, R., private communication.  More information about Y4KCam is available online at http://astronomy.ohio-state.edu/Y4KCAM
\end{thebibliography}
\end{document}